\documentclass[11pt,a4paper]{article}
\usepackage[utf8x]{inputenc}
\usepackage[pdftex]{graphicx}
\usepackage[margin=1.3in]{geometry}

\usepackage{algorithm2e}
\usepackage{amsthm}
\usepackage{amsmath}
\mathcode`.="613A          % . as punctuation
\mathcode`|="326A          % | as a relation

\def\land{\mathrel{\wedge}}

\def\preleq{\mathrel{%      % pre-order leq
  \raise 2pt\hbox{$\mathop\sqsubset\limits_{\hbox{$\sim$}}$}%
}}

\let\union=\cup
\let\bigunion=\bigcup

\newcommand\setsize[1]{\left| #1 \right|}

\newcommand\auxfun[1]{\expandafter\newcommand\csname #1\endcsname{%
 \mathop{\hbox{\rm #1}}\nolimits}}

\newcommand\af[1]{\mathop{\hbox{\rm #1}}\nolimits}

\auxfun{Max}
\auxfun{fix}
\auxfun{dom}
\auxfun{ran}

\title{Fast Distributed Computation of Distances in Networks}

\newtheorem{lemma}{Lemma}[section]
\newtheorem{definition}{Definition}[section]
\newtheorem{proposition}{Proposition}[section]
\newtheorem{theorem}{Theorem}[section]
\newtheorem{corollary}{Corollary}[section]

\newcommand\dist{\af{d}}
\auxfun{ecc}
\auxfun{diam}
\auxfun{radius}

\begin{document}

%\author{
%\IEEEauthorblockN{Paulo Sérgio Almeida}
%\IEEEauthorblockA{DI/CCTC\\
%Universidade do Minho\\
%Braga, Portugal\\
%psa@di.uminho.pt}
%\and
%\IEEEauthorblockN{Carlos Baquero}
%\IEEEauthorblockA{DI/CCTC\\
%Universidade do Minho\\
%Braga, Portugal\\
%cbm@di.uminho.pt}
%\and
%\IEEEauthorblockN{Alcino Cunha}
%\IEEEauthorblockA{DI/CCTC\\
%Universidade do Minho\\
%Braga, Portugal\\
%alcino@di.uminho.pt}
%}

\author{Paulo S\'{e}rgio Almeida \and Carlos Baquero \and Alcino Cunha}
\date{HASLab / INESC TEC, Universidade do Minho, Braga, Portugal}

\maketitle

\begin{abstract}
This paper presents a distributed algorithm to simultaneously compute the
diameter, radius and node eccentricity in all nodes of a synchronous
network.  Such topological information may be useful as input to configure
other algorithms.
Previous approaches have been modular, progressing in sequential phases
using building blocks such as BFS tree construction, thus incurring 
longer executions than strictly required.
We present an algorithm that, by timely propagation of available
estimations, achieves a faster convergence to the correct values.  We show
local criteria for detecting convergence in each node.  The algorithm avoids
the creation of BFS trees and simply manipulates sets of node ids and hop
counts.
For the worst scenario of variable start times, each node $i$
with eccentricity $\ecc(i)$ can compute: the node eccentricity in
$\diam(G)+\ecc(i)+2$ rounds; the diameter in $2\diam(G)+\ecc(i)+2$ rounds;
and the radius in $\diam(G)+\ecc(i)+2\radius(G)$ rounds.

\end{abstract}

\section{Introduction}

This paper presents a distributed algorithm to simultaneously compute the
diameter $D$, radius $R$ and node eccentricity $\ecc(i)$ in all nodes of a
network.  An early knowledge of this topological information is useful since
it is often used as input to other algorithms. For instance, the diameter or
eccentricity can be used to simplify termination in leader election
algorithms \cite{DBLP:books/mk/Lynch96} and calibrate \emph{time-to-live}
parameters \cite{DBLP:journals/ijnm/LeeBP03}; the radius and eccentricity
allow determining center nodes \cite{DBLP:journals/toplas/KorachRS84}, which
are nice candidates to serve as coordinators in other distributed
algorithms.

We assume a synchronous network model, while allowing variable start times,
in which one or more nodes can start the algorithm with no prior
coordination.  The algorithm is designed to be fast in a precise sense; we
are concerned with, not just asymptotic complexity, but exact bounds in the
number of rounds.

The classic approach to this problem \cite{DBLP:books/mk/Lynch96} is to
compute the eccentricities by parallel construction of \emph{breadth first
search} (BFS) trees 
%\cite{AG85}) 
rooted at each node.  Once eccentricities are known, each BFS
tree can be reused to do a global computation, starting from the leafs and
converging to each root node, allowing each to compute the maximum and
minimum eccentricity (the network diameter $D$ and radius $R$).  Considering
a graph $G = (V,E)$, this classic approach has total message complexity of
$\Theta(\setsize V \setsize E \log \setsize V)$ bits.  The diameter and
radius are known at all nodes in at most $4 D +2$ rounds. 

These time bounds can be improved if one departs from this modular
multi-phase approach, where BFS trees are first constructed to compute
eccentricities. This paper introduces an algorithm that propagates candidate
values in a timely and continuous fashion, resulting in a faster convergence
to the correct values.
The challenge in this strategy is that a suitable termination method must be
devised to detect, in each node, when the candidate values 
have converged.  Under the same message complexity of the classic approach, the
proposed algorithm reduces the number of rounds to compute the diameter to
at most $3D+1$ rounds and the radius to at most $2D+2R$ rounds.
To be more precise, with this algorithm each node $i$ with
eccentricity $\ecc(i)$ computes:

\begin{itemize}
\item the node eccentricity at most by round $\diam(G) + \ecc(i) +2$; 
\item the diameter at most by round $2 \diam(G) + \ecc(i) +2$; and
\item the radius at most by round $\diam(G) + \ecc(i) + 2 \radius(G)$. 
\end{itemize}

The paper is organized as follows. Section \ref{sec:model} presents the
computing model and introduces notation. The algorithm is presented in
Section \ref{sec:algorithm}. Example runs of the algorithm, proofs of local
convergence criteria and global convergence bounds are also included in this
section.  The related work is discussed in Section \ref{sec:related-work},
and the conclusions are presented in Section \ref{sec:conclusions}.

\section{Network Model and Notation}
\label{sec:model}

We assume a synchronous network model, similar to the one described
in~\cite{DBLP:books/mk/Lynch96}.  The network is composed by a set of nodes
connected by links, which we assume to be bidirectional; i.e., we have a
simple, connected, unweighted, undirected graph $G = (V, E)$
with $\setsize V \geq 2$ nodes and $\setsize E \geq 1$ links. We assume
globally unique identifiers for nodes, but no knowledge of the
network topology or the number of nodes.

Computation proceeds in synchronous rounds. At each round, nodes first look
at their state and compute what messages are sent, through a
\emph{message-generation function}; then nodes look at their state and
messages received and compute the new state after the round, through a
\emph{state-transition function}.  We assume no link or process failures.
In order to obtain an asynchronous version of the algorithm a synchronizer
$\alpha$ \cite{DBLP:journals/jacm/Awerbuch85} can be used. 
We operate under a maximum bandwidth of $O(\setsize V \log \setsize V)$
bits, per link per round. 

We assume the general case with variable start times. Nodes start as
\emph{quiescent}, a state in which they do not send messages nor transition
to different states. Nodes wakeup when they receive a message from a special
\emph{environment node} (not part of $G$, connected to every node), or from
an already \emph{active} node.

We use $\dist(i, j)$ to denote the distance between nodes $i$ and $j$ (the length
of the shortest path between nodes $i$ and $j$); $\ecc(i)$ for the eccentricity of
node $i$ (the maximum $\dist(i,j)$ between node $i$ and any other node $j$);
$\diam(G)$ for the network diameter (the maximum eccentricity over all
nodes); $\radius(G)$ for the network radius (the minimum eccentricity
over all nodes); and $\af{nbrs}(i)$ for the set of neighbors of node $i$
(nodes connected to node $i$ by a link).

\section{Algorithm}
\label{sec:algorithm}

\newcommand\nn{c}
\newcommand\la{\langle}
\newcommand\ra{\rangle}
\newcommand\const{\textsc}
\newcommand\bfstup[1]{\la \const{bfs}, #1 \ra}
\newcommand\diamtup[1]{\la \const{diam}, #1 \ra}
\newcommand\radtup[1]{\la \const{rad}, #1 \ra}

The algorithm is presented in Figure~\ref{algorithm}. At each round, a node $i$
sends the same message to all its neighbors (state variable $O_i$). A
message is a non-empty set of tuples; the empty set represents absence of a
message.
The tuples in a message can be $\bfstup{\_,\_}$, $\diamtup{\_}$ or
$\radtup{\_}$, where \const{bfs}, \const{diam} and \const{rad} are
constants.
Nodes do not need to distinguish between messages that arrive from
different neighbors; the second parameter of the state-transition function
(parameter $M_i$) is the set of messages received by node $i$ from all 
neighbors.

\begin{figure}
\hspace{-6mm}
\begin{minipage}{1.2\hsize}
%\restylealgo{boxed}
\begin{algorithm}[H]
\SetKwBlock{OnConst}{constants:}{}
\SetKwBlock{OnState}{state variables:}{}
\SetKwBlock{OnStateTrans}{state-transition function:}{}
\SetKwBlock{OnMsgGen}{message-generation function:}{}
\SetVline
\OnState{
	$e_i$, node eccentricity, initially $e_i=0$\\
	$d_i$, network diameter, initially $d_i=0$\\
	$r_i$, network radius, initially $r_i=\infty$\\
	$s_i$, status, initially $s_i=\const{quiescent}$\\
	$I_i$, set of node ids, initially $I_i=\{\}$\\
	$\nn_i$, consecutive rounds with no new BFS, initially $\nn_i=0$\\
	$O_i$, message to be sent, initially $O_i=\{\}$\\
}
\BlankLine
\OnMsgGen{
$\af{msg}_{i}(\la e_i, d_i, r_i, s_i, I_i, \nn_i, O_i \ra,j) = O_i
  \qquad j \in \af{nbrs}(i)$
%\tcp{Sends same msg to all neighbours}
}
\OnStateTrans{
$\af{trans}_{i}(\la e_i, d_i, r_i, s_i, I_i, \nn_i, O_i \ra, M_i) = 
        \la e_i', d_i', r_i', s_i', I_i', \nn_i', O_i' \ra $
\BlankLine
\textbf{where}
\BlankLine
%$M=\bigunion \{ M_{ji} |  M_{ji} \in M_i \}$\\
$M=\bigunion \{ m |  m \in M_i \}$\\
\eIf{$s_i = \const{quiescent} \land M = \{\}$} {
  $ \la e_i', d_i', r_i', s_i', I_i', \nn_i', O_i' \ra =
    \la e_i, d_i, r_i, s_i, I_i, \nn_i, O_i \ra $
}{
  $M' = \{ \bfstup{j, h+1} | \bfstup{j, h} \in M, j \not\in I_i \}$ \\
  $M'' = M' \union \{ \bfstup{i, 0} | s_i = \const{quiescent} \}$\\
%  \eIf{$M' = \{\} \land s_i = \const{active}$} {
  \eIf{$M'' = \{\} $} {
    $\nn_i' = \nn_i +1$\\
  }{
    $\nn_i' = 0$\\
  }
  $e_i' = \max(\{e_i\} \union \{ h | \bfstup{\_, h} \in M' \})$\\
  $d_i' = \max(\{d_i\} \union \{ d | \diamtup{d} \in M \} \union \{e_i'\})$\\
  $r_i' = \min(\{r_i\}
          \union \{ r | \radtup{r} \in M \}
          \union \{ e_i' | \nn_i' = 2 \})$\\
  $s_i' = \const{active}$ \\
  $I_i' = I_i \union \{ j | \bfstup{j, \_} \in M'' \} $ \\
  $M^d = \{ \diamtup{d_i'} | d_i' > d_i\}$\\
  $M^r = \{ \radtup{r_i'} | r_i' < r_i\}$\\
  $O_i' = M'' \union M^d \union M^r$
%  \eIf{$s_i = \const{quiescent}$} {
%    $M'' = M' \union \{ \bfstup{i, 0} \}$
%  }{
%    $M'' = M'$
%  }
%  \eIf{$d_i' > d_i$} {
%    $M''' = M'' \union \{ \diamtup{d_i'} \}$
%  }{
%    $M''' = M''$
%  }
%  \eIf{$r_i' < r_i$} {
%    $O_i' = M''' \union \{ \radtup{r_i'} \}$
%  }{
%    $O_i' = M'''$
%  }
}
}
\end{algorithm}
\end{minipage}
\caption{Algorithm.}
\label{algorithm}
\end{figure}

Each node when awoken broadcasts a \const{bfs} message with its id and a hop
counter, which starts at 0. Nodes keep the set of ids of all received
\const{bfs} messages.  When a node receives a \const{bfs} message from a
node not yet known, it increments the hop counter and rebroadcasts it.

Nodes know their eccentricity is at least the largest hop count received in
a \const{bfs} message, which they keep in a variable ($e_i)$. Nodes also
keep in two other variables ($d_i$, $r_i$) a lower bound estimate for the
diameter and an upper bound estimate for the radius. When a node increases
the diameter estimate; it broadcasts a \const{diam}  message with the new
value. A node increases its diameter value when (1) its eccentricity
surpasses its diameter estimate or (2) it receives a \const{diam}
message whose value is higher that its diameter value.  Estimation of the
radius is also driven by eccentricity, but must be deferred until each node
detects its correct eccentricity. When that happens, and also when a lower
estimate for the radius is received, a \const{rad} message is broadcast. 

It is easy to see that at some point every node will have awakened; later
everyone will have received a \const{bfs} from everyone else and will have
their eccentricity stored in the respective variable; and later still nodes
will receive some \const{diam} message with the network diameter,
originating from some maximum eccentricity node (a periphery node).
Similarly, a \const{rad} message originating from a minimum eccentricity
node (a center node) will arrive eventually at all nodes.

The relevant question is convergence detection, i.e., when will nodes know
that their eccentricity, diameter and radius variables have converged to the
correct values. For this purpose, and inspired by the approach in
\cite{DBLP:journals/jpdc/Peleg90}, nodes have a variable which stores the
number of consecutive rounds for which no new \const{bfs} messages arrived.
Later, we will show how this variable can be used for convergence detection.

In order to analyze the communication complexity of the whole execution of
the algorithm, we can first observe that each \const{bfs} message
can be encoded in $\Theta(\log \setsize V)$ bits, since its
dominated by the size of the ids. Each node retransmits
exactly one \const{bfs} message for every other node, totaling
$\Theta(\setsize V \setsize E \log \setsize V)$ bits. 
Since the diameter can only increase at most $D$ times, each node broadcasts
at most $D$ \const{diam} messages, totaling $O(D \setsize E \log D)$ bits.
Similarly for \const{rad} messages. Thus, total message complexity is
$\Theta(\setsize V \setsize E \log \setsize V)$ bits.

\subsection{Example Runs}

\begin{figure}
\center{
\includegraphics[ scale=0.4 ]{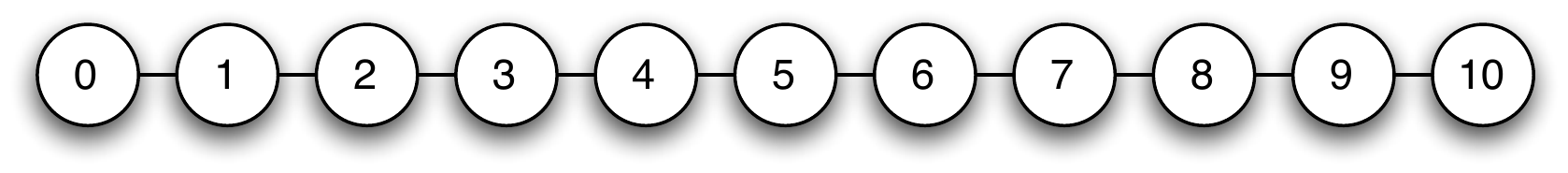}
}
\caption{Path graph.}\label{fig:pathg11}
\end{figure}

\begin{figure}
\center{
\includegraphics[ scale=0.7 ]{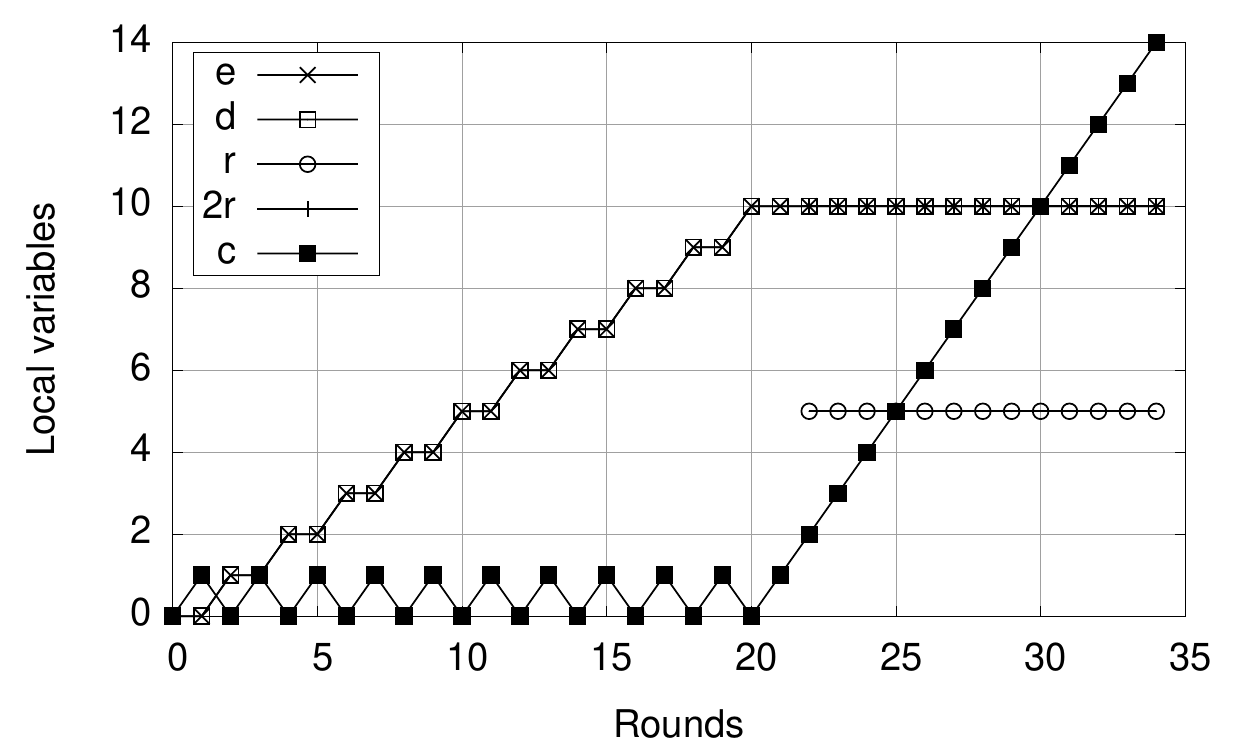}
}
\caption{Path graph, start at 0, probing at node 0.}\label{fig:rung11p0}
\end{figure}

\begin{figure}
\center{
\includegraphics[ scale=0.7 ]{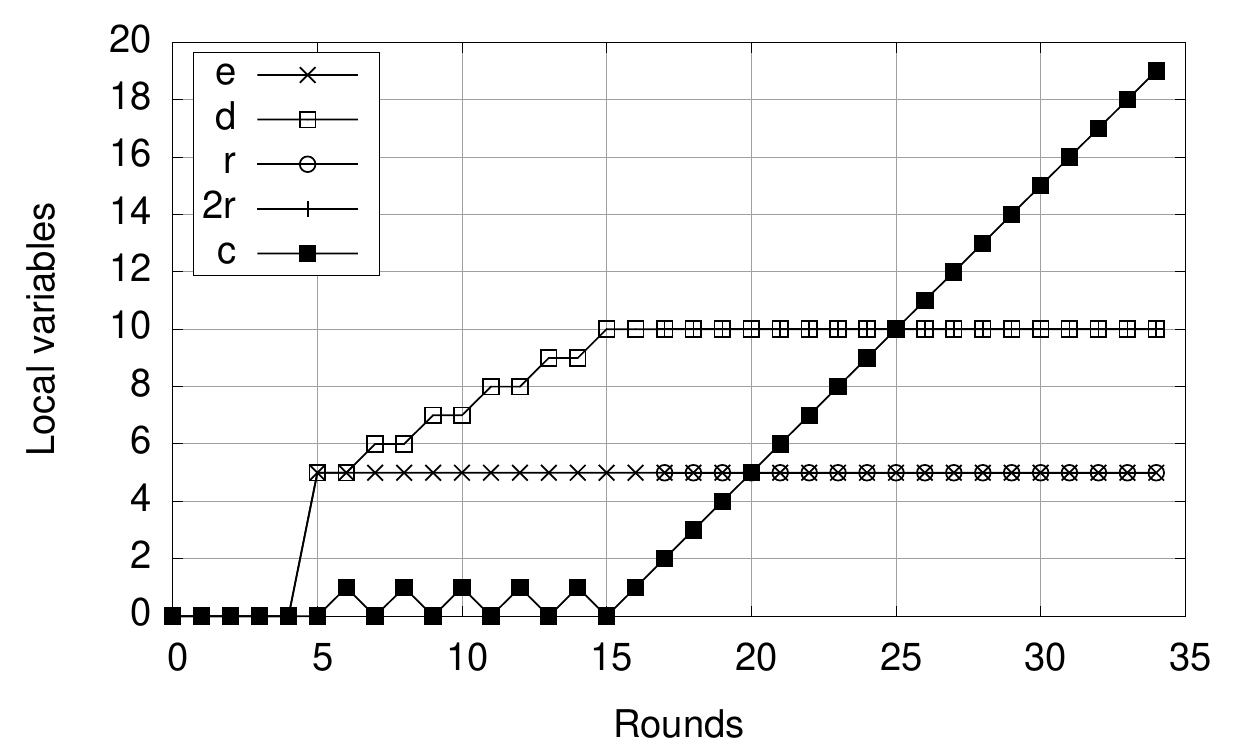}
}
\caption{Path graph, start at 0, probing at node 5.}\label{fig:rung11p5}
\end{figure}

\begin{figure}
\center{
\includegraphics[ scale=0.7 ]{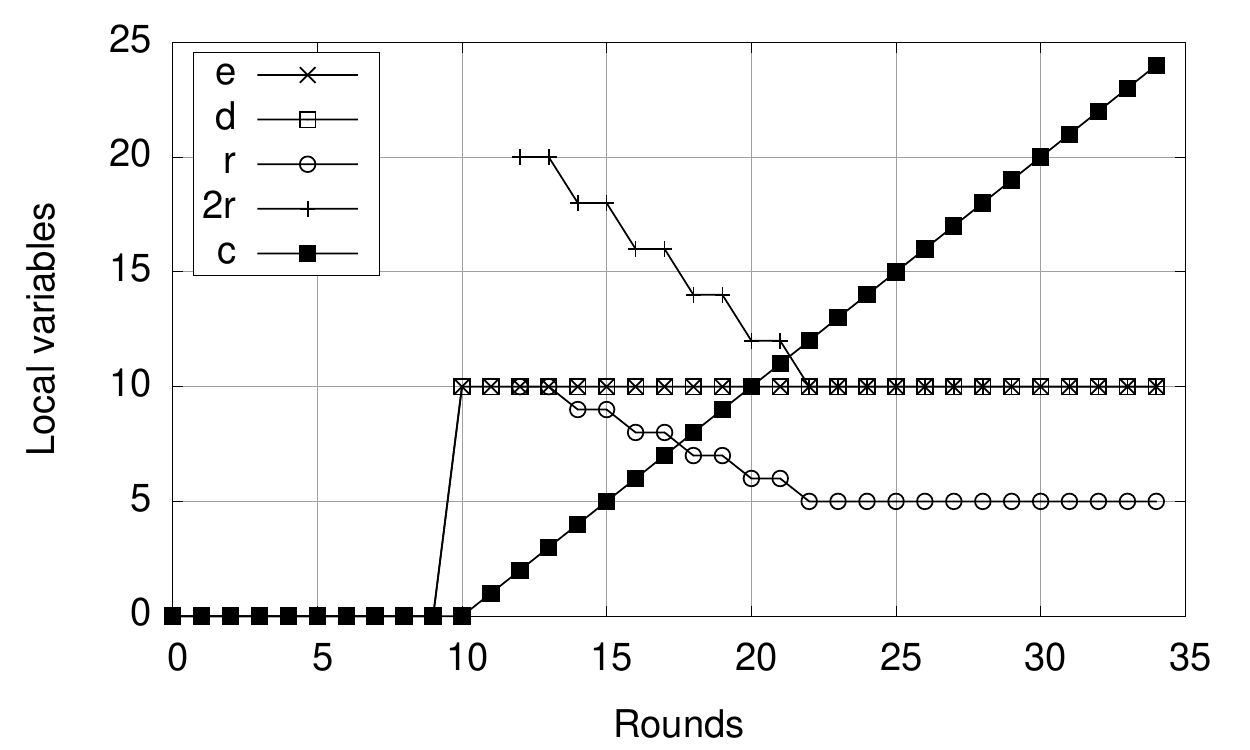}
}
\caption{Path graph, start at 0, probing at node 10.}\label{fig:rung11p10}
\end{figure}

Prior to the formal proofs of the algorithm properties, we now
convey some intuition by illustrating its execution in two different
graphs.
The first graph is a path with eleven nodes, depicted in Figure
\ref{fig:pathg11}. Nodes 0 and 10 and are the only two nodes in the
periphery, thus defining the diameter; node 5 is the single node in the graph
center.  We
consider a run where node 0 is activated and we will observe how the local
variables evolve in nodes 0, 5 and 10 (Figures \ref{fig:rung11p0},
\ref{fig:rung11p5} and \ref{fig:rung11p10}, respectively). 

When probing the variables at node 0 we can observe that each two
rounds a new \const{bfs} is
received until \const{bfs} messages from all nodes have arrived. Each time a new \const{bfs} is
received the
$\nn_i$ counter is reset.
Thus, when $\nn_i$ reaches two the node knows that
its eccentricity variable $e_i$ has reached the correct final value. This
happens in nodes 0, 5 and 10 at rounds 22, 17 and 22, respectively. 

Once eccentricity is stabilized nodes start disseminating \const{rad} messages in
order to detect the minimum eccentricity. Since local radius estimates can
only decrease, one must establish a termination condition so that each node
knows that it reached the correct radius. We will show that this is safely
achieved when $\nn_i \geq 2 r_i$. The run in Figure \ref{fig:rung11p10} shows
that no sooner than this has the radius reached its final value of 5 in
node 10. 

Diameter dissemination, via \const{diam} messages, starts even if nodes are still
updating their (monotonically increasing) estimates for eccentricity. 
For instance, in Figure
\ref{fig:rung11p5} we see that node 5 is activated at round 5 (receiving
\const{bfs} messages
from nodes 0 to 4) and sets both $e_i$ and $d_i$ to 5 (its received hop
distance from node 0, the furthest away). As nodes 6 to 10 are activated, in
the next rounds, the diameter estimate $d_i$ at node 5 gets updated each two
rounds. 
We will show that in order for a node to know that the diameter has reached
its final value it needs to locally observe that $\nn_i \geq 2$ and $\nn_i >
d_i$.  This happens in nodes 0, 5 and 10 at rounds 31, 26 and 21,
respectively. One can notice that in this path graph the diameter is
stable even before those rounds. 

\begin{figure}
\center{
\includegraphics[ scale=0.4 ]{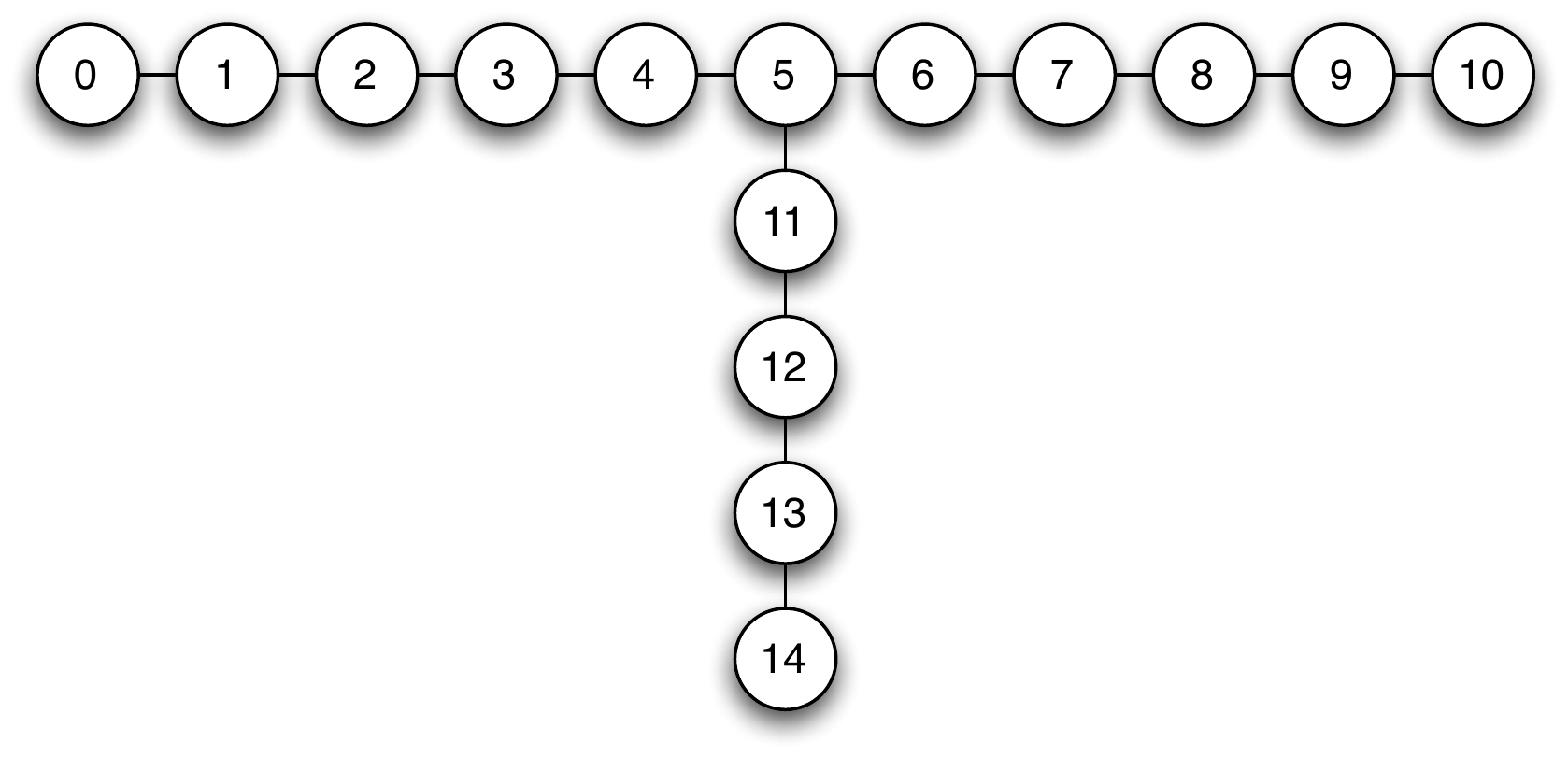}
}
\caption{T shaped graph.}\label{fig:tg53}
\end{figure}

\begin{figure}
\center{
\includegraphics[ scale=0.7 ]{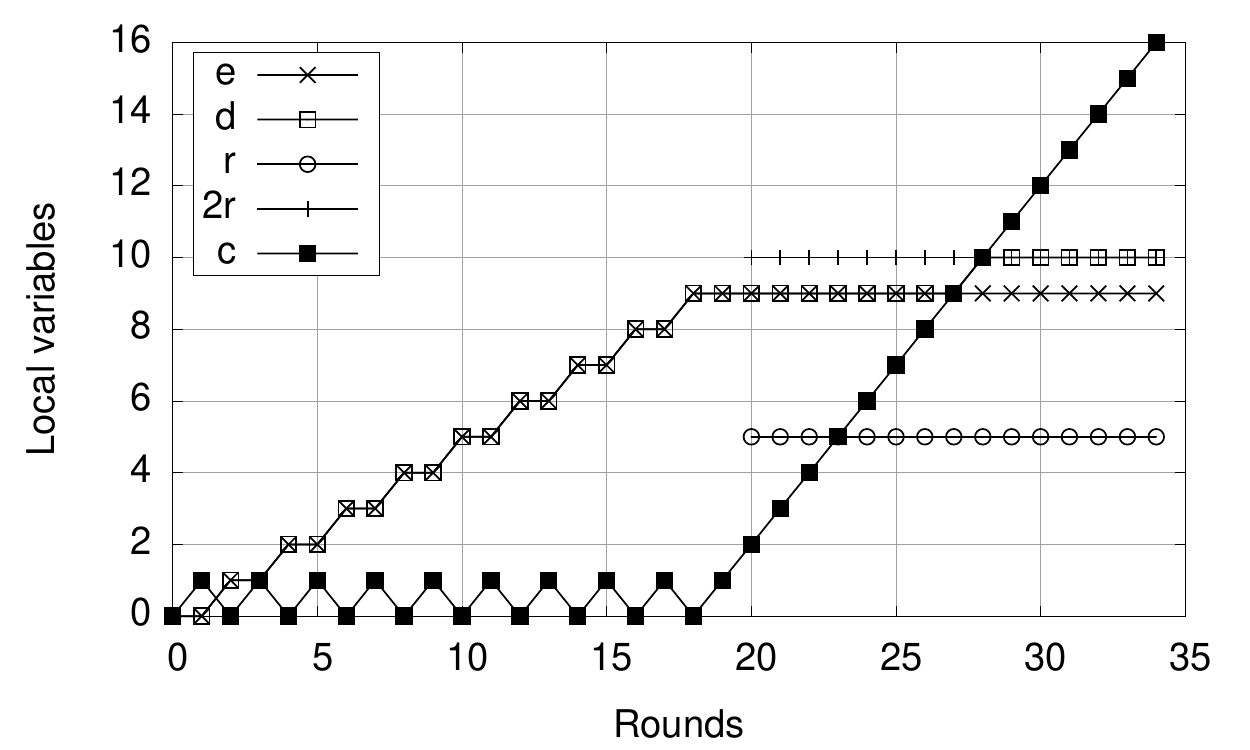}
}
\caption{T shaped graph, start at 14, probing at node 14.}\label{fig:rung53p53}
\end{figure}

However, it is easy to construct a T shaped graph where detection cannot
occur before the mentioned round.  In Figure \ref{fig:tg53} we show such a
graph. Here, a critical case happens when a node not in a diameter defining
path (for example, node 14) is the first to be activated. As seen in
Figure~\ref{fig:rung53p53} this leads to a
run where the diameter estimation seems to be stable for a large number of
rounds (between round 18 and 27), finally increases at round 28, with 
convergence being detected only when $\nn_i > d_i$ 
at round 29.  
In the next section we prove that these convergence criteria are
correct for general graphs.

\subsection{Local Convergence Criteria}

We now establish results that allow a node to know, using local information,
that the variables estimating node eccentricity, network diameter and
network radius have converged to the correct values.
In the following we will use $i$, $j$, $u$, and $v$ to range over node ids and
$r$, $n$ and $k$ to range over non-negative integers.

\begin{definition}
$
A_i(r) \doteq \{j | \dist(i,j) \leq r \land s_j(r-\dist(i,j))=\const{active}\}.
$
\end{definition}

$A_i(r)$ denotes the ``area of visibility'' of node $i$ after%
\footnote{When referring to state variables, we do not use the expression ``at
round $r$'' to avoid ambiguity between beginning or end of round.
Throughout the paper we use ``after round $r$'' as a shorthand for
``when round $r$ has finished'', i.e., ``at the end of round $r$''.}
round $r$: the
set of nodes whose \const{bfs} messages arrive at $i$ no later than round $r$.
It is easy to see that $A_i$ is monotonic: $A_i(r) \subseteq A_i(r+1)$.

\begin{lemma}
After any round $r$, $I_i(r) = A_i(r)$. 
\end{lemma}
\begin{proof}
If $j \in I_i(r)$ then $i$ must have received no later than round $r$ a
\const{bfs}
message starting at $j$; it traveled $\dist(i,j)$ hops, which means that
$\dist(i,j) \leq r$ and $j$
was active after round $r - \dist(i,j)$; therefore $j \in A_i(r)$.
If $j \in A_i(r)$, then $\dist(i,j) \leq r$ and $j$ was active after round
$r - \dist(i,j)$. This implies that $j$'s \const{bfs} message arrives at $i$ not later
than round $r$; therefore, $j \in I_i(r)$.
\end{proof}

\begin{lemma}
	\label{lem:nnav}
If $\nn_i(r+2) \geq 2$, then $A_i(r) = V$.
\end{lemma}
\begin{proof}
If $A_i(r) \neq V$ then there are two nodes, $u \in A_i(r)$ and $v
\not \in A_i(r)$ which are adjacent, i.e., $\dist(u,v)=1$.
This means that
$s_u(r-\dist(i,u)) = \const{active}$
and
$s_v(r-\dist(i,v)) = \const{quiescent}$.
Since $u$ and $v$ are adjacent, then $s_v(r+1-d(i,u)) = \const{active}$.
There are three possible cases:
(1) $\dist(i,u) = \dist(i,v)$, in which case
$v$ became active in round $r+1-\dist(i,v)$, and $i$ receives the
\const{bfs}
from $v$ at round $r+1$;
(2) $\dist(i,v) = \dist(i,u) + 1$, in which case $s_v(r+2-\dist(i,v)) =
\const{active}$, and $i$ receives the \const{bfs} from $v$ at either round $r+1$ or
round $r+2$;
(3) $\dist(i,v) = \dist(i,u) - 1$, cannot happen, as it contradicts
$s_v(r-\dist(i,v)) = \const{quiescent}$.
In any case, $\nn_i(r+1) = 0$ or $\nn_i(r+2) = 0$.
Therefore, since $\nn_i$ can only increase 1 unit per round,
it follows that if $\nn_i(r+2) \geq 2$, then $A_i(r) = V$.
\end{proof}

\begin{lemma}
After any round $r$,
$e_i(r) = \max(\{0\} \union \{\dist(i,j) | j \in I_i(r) \})$.
\end{lemma}
\begin{proof}
Trivial induction on the number of rounds.
\end{proof}

\begin{theorem}[eccentricity convergence]
	\label{the:nne}
If $\nn_i(r+2) \geq 2$, then $e_i(r) = \ecc(i)$.
\end{theorem}
\begin{proof}
Combine the previous three lemmas.
\end{proof}

\begin{lemma}
	\label{lem:nnincr}
If for some $r$ and $n \geq 2$, $\nn_i(r+n) = n$, then for all $k$, $\nn_i(r+k) = k$.
\end{lemma}
\begin{proof}
After the first round $r$ such that  $\nn_i(r+2) = 2$, by the first two lemmas
$I_i(r) = V$. This implies that for $r' \geq r$, $\nn_i(r'+1) = \nn_i(r') +
1$.
\end{proof}

\begin{theorem}[diameter convergence]
	\label{the:ldiam}
When $\nn_i(r) \geq 2$ and $\nn_i(r) > d_i(r)$, then $d_i(r) = \diam(G)$.
\end{theorem}

\begin{proof}
By contradiction.
Assume $\nn_i(r) \geq 2$ and $\nn_i(r) > d_i(r)$ but $d_i(r) < \diam(G)$.
By Theorem~\ref{the:nne}, $e_i(r) = \ecc(i)$.
Also, it is trivial that $d_i(r) \geq e_i(r)$.
Then, as in a graph all the eccentricities between the radius and the
diameter are present \cite{BH90}, there exists two nodes $u$ and $v$ with $\dist(u,v) =
d_i(r) + 1$. Assume without loss of generality that $\dist(i,u) \geq
\dist(i,v)$.
From the previous lemma, for $r' = r - \nn_i(r)$, it follows that
$\nn_i(r'+k) = k$.
As $\nn_i(r'+2) = 2$, by Lemma~\ref{lem:nnav}, $A_i(r') = V$;
therefore $s_u(r' - \dist(i,u)) = \const{active}$.
Then,
$d_v(r'-\dist(i,u)+\dist(u,v)) \geq \dist(u,v)$
(\const{bfs} from $u$ has reached $v$). Furthermore,
$d_i(r'-\dist(i,u)+\dist(u,v)+\dist(i,v)) \geq \dist(u,v)$
(\const{diam} message from $v$ has reached $i$).
Since $\dist(i,u) \geq \dist(i,v)$, then
$d_i(r'+\dist(u,v)) \geq \dist(u,v)$.
Recall that $\dist(u,v) = d_i(r) + 1$, and let
$r'' = r' +\dist(u,v) = r - \nn_i(r) + d_i(r) + 1$.
From the assumption $\nn_i(r) > d_i(r)$, it means that $r'' \leq r$,
which together with $d_i(r'') \geq d_i(r) + 1$ contradicts the monotonicity
of $d_i$.
\end{proof}

\begin{theorem}[radius convergence]
When $\nn_i(r) \geq 2r_i(r)$, then $r_i(r) = \radius(G)$.
\end{theorem}

\begin{proof}
%For a single node network, the first assignment of $r_i$, from $\infty$ to
%0, happens when $\nn_i = 2$.
We assume networks with at least one link and two nodes, which
means $r_i(r) \geq 1$. If $\nn_i(r) \geq 2r_i(r)$, we have $\nn_i(r) \geq
2$, which means that, from Lemma~\ref{lem:nnincr}, all \const{bfs}s have already
reached 
node $i$ after round $r' = r - \nn_i(r)$, and from $r'$ on we have
$\nn_i(r'+k) = k$. Assume, by contradiction, that $r_i(r) > 
\radius(G)$. Then, at most after round $r' + r_i(r) - 1$, all
\const{bfs}s have
reached some node $u$ in the center of the network. At most two rounds
later, after round $r''= r' + r_i(r) + 1$, we have $\nn_u(r'') \geq 2$ and
$u$ has sent a \const{rad} message with the network radius. At most $r_i(r) - 1$
rounds later, after round $r''' = r' + 2r_i(r)$ this message arrives at $i$ and $r_i(r''') =
\radius(G)$. But assuming $\nn_i(r) \geq 2r_i(r)$ it means that $r''' \leq
r$, which contradicts $r_i$ being monotonically decreasing.
As $r_i$ results from some eccentricity and is always an upper bound of the
radius, we must have $r_i(r) = \radius(G)$.
\end{proof}

\subsection{Convergence Bounds}

We now determine upper bounds on the number of rounds for convergence of
eccentricity, diameter and radius.
Given that we have described the algorithm for the general case of variable
starting times, what matters is the number of rounds after the first
activation; i.e., ignoring an initial sequence of rounds with all nodes
inactive. Therefore, in this section we consider that the first node became
active after round 0; round 1 is when the first non-environment message is
sent.

\begin{proposition}[eccentricity bound]
Node $i$ can determine its eccentricity at most in
$\diam(G) + \ecc(i) + 2$ rounds.
\end{proposition}

\begin{proof}
After round $\diam(G)$ all nodes are active, so the last $\const{bfs}$ arrives
at $i$ at most after round $\diam(G) + \ecc(i)$. Two rounds later the $\nn_i$
variable reaches $2$ and from Lemma~\ref{the:nne} the eccentricity has
already converged.
\end{proof}

\begin{proposition}[diameter bound]
\label{pro:diam}
Node $i$ can determine the network diameter at most in
$2 \diam(G) + \ecc(i)  +1$ rounds.

\end{proposition}
\begin{proof}
After round $\diam(G)$ all nodes are active, so the last $\const{bfs}$ arrives
at $i$ at most after round $\diam(G)+ \ecc(i)$. Subsequently $\nn_i$ starts
increasing and after further $\diam(G) +1$ rounds the local condition $\nn_i >
d_i$ is met. As we are considering networks with at least one link, i.e.,
$\diam(G) \geq 1$, then at this round we have also $\nn_i \geq 2$ and from
Theorem~\ref{the:ldiam} the diameter has converged.
\end{proof}

\begin{proposition}[radius bound]
\label{pro:radius}
Node $i$ can determine the network radius at most in
$\diam(G) + \ecc(i)  + 2 \radius(G)$ rounds.
\end{proposition}
\begin{proof}
After round $\diam(G)$ all nodes are active, so the last $\const{bfs}$ arrives
at $i$ at most after round $\diam(G)+ \ecc(i)$; afterwards $\nn_i$ starts
increasing and after round
$r = \diam(G) + \ecc(i) + 2 \radius(G)$ we have $\nn_i(r) \geq 2\radius(G)$.
Also, after at most round $\diam(G) + \radius(G)$ all $\const{bfs}$ have
arrived at all center nodes; two rounds later, at most after round
$r' = \diam(G) + \radius(G) + 2$, each center node sends a
$\const{rad}$ message containing $\radius(G)$.
There are two possibilities: (1) $\ecc(i) > 1$, the
$\const{rad}$ message from a center node $j$ arrives at $i$ at most
$\radius(G)$ rounds later, which means that at most after
round $r''= \diam(G) + 2\radius(G) + 2 $ we have $r_i(r'') = \radius(G)$;
given that $r'' \leq r$, then $r_i(r) = \radius(G)$, and the local radius
convergence criteria $\nn_i(r) \geq 2r_i(r)$ is met; (2) $\ecc(i) = 1$,
in which case $\radius(G) = 1$,  $i$ is a center node, and at round
$r'$ we have $r_i(r') = \radius(G)$; as in this case $r' = r$, we have
$\nn_i(r) \geq 2r_i(r)$ as well.
\end{proof}

\begin{corollary}
\label{cor}
All nodes know: their eccentricity at most in $2D+2$ rounds; the diameter at
most in $3D+1$ rounds; and the radius at most in $2D+2R$ rounds.
\end{corollary}

\subsection{Termination}

To keep the presentation clear and avoid cluttering, we did not include in
the algorithm the mechanics of termination; i.e., each node reaching a
``terminated'' state in which it stops sending messages. In general
distributed termination is independent from reaching some result, and nodes
may have to keep propagating messages for some time.

In this case, however, it is easy to see  that when a node has determined
both the radius and diameter through the local convergence criteria, all
neighbors will have the same criteria met after at most one more round.
(After one more round, each node $j$ neighbor from $i$, will have $c_j$
with at least the same value node $c_i$ had, and both $r_j$ and $d_j$ will
have the same values as in node $i$.)
Therefore, after having met both criteria for radius and diameter, a node
needs only execute one more round and stop.

\subsection{Improving Storage Requirements}

In the previously described algorithm each node accumulates in the $I$
variable all ids received in all previous rounds. Although it has made the
description intuitive and streamlined proofs, it means that, regardless of
network topology, by the end of the execution each node will need to store
$\Theta(\setsize V)$ ids.

Here we show that it is enough to keep in the state only the ids received in
the two previous rounds. While this modification does not change the worst
case space requirement complexity (it still remains $O(\setsize V)$ ids for
general graphs and uncoordinated start times), it may be useful in practice.
As an example, for 2D geometrical networks (e.g. a geographically spread
sensor network with links according to inter-node distance), under
synchronized start times, the number of ids that arrive in a single round
(and need to be stored) will be $O(\sqrt{\setsize V} )$. Notice that these
specific configurations also reduce the required channel bandwidth, and that
in other specific graph topologies, these uppers bounds on stored state can
be even more tight. 

The modification to the algorithm is trivial and consists of replacing
state variable $I$ by a pair $I, J$ used as a sliding window;
replacing the test $j \not\in I_i$ with $j \not\in I_i \union J_i$ and
replacing $I_i' = I_i \union \{ j | \bfstup{j, \_} \in M'' \} $ with
$I_i' = J_i$ and $J_i' = \{ j | \bfstup{j, \_} \in M'' \}$. The modification
is possible due to the following property of the original algorithm.

\begin{proposition}
A node $j$ can only receive $\const{BFS}$ messages $\bfstup{i, \_}$,
originated in a node $i$ activated at round $r$, in rounds $r + \dist(i,j)$,
$r + \dist(i,j) + 1$ and $r + \dist(i,j) + 2$.
\end{proposition}
\begin{proof}
The first round, where node $j$ can receive a $\bfstup{i,\_}$ message, is $r' = r + \dist(i,j)$, by the shortest path from $i$; $j$ rebroadcasts it and at round
$r' + 1$ it arrives at all neighbors, if any.
Also at round $r'+1$ node $j$ may receive such a message if there exists a
neighbor node $u$ at the same distance from $i$ (i.e., $\dist(i,u) =
\dist(i,j)$); this neighbor, similarly to $j$, has received such a message
at round $r'$ and has rebroadcasted it.
At round $r'+2$ node $j$ can receive such a message if it has a neighbor
$u$ one hop further away from $i$ (i.e., $\dist(i,u) = \dist(i,j) + 1$); this
neighbor has received the message at round $r'+1$ and has rebroadcasted it
at round $r'+2$.
Because any neighbor $u$ of $j$ has stored $i$ in the respective $I_u$
variable after at most round $r'+1$, it will not rebroadcast any
$\bfstup{i,\_}$ message in any round later than $r'+2$, from which such
messages cannot reach $j$ later than round $r'+2$.
\end{proof}

%This simple technique can have a broather application towards enacting state
%savings in other syhnchronous flooding algorithms. 

\section{Related Work}
\label{sec:related-work}

As mentioned in the introduction, our algorithm improves the classic modular
approach~\cite{DBLP:books/mk/Lynch96}, where BFS trees are first computed at
each node and later reused to perform a global computation of the network
radius and diameter (in at most $4D+2$ rounds). By Corollary~\ref{cor}, we
achieve a speedup of $D$ rounds for computing the diameter, and since $D \le
2R \le 2D$ the speedup for computing the radius varies between $2$ and $D+2$
rounds. The maximum speedup occurs, for example, in path graphs.
This improvement is achieved with the same message
complexity of $\Theta(\setsize V \setsize E \log
\setsize V)$ bits, and the same space complexity of $O(\setsize V)$ ids 
and computation complexity per round of
$O({\setsize V}^2)$ at each node.

The work in \cite{DBLP:conf/podc/SzymanskiSP85} computes the diameter under
the more restrictive synchronized start time model, where all nodes are
activated at the first round. This is a fast algorithm since they also
disseminate candidate values for eccentricities before they
converge. However, since they assume that all nodes are
active in the first round the local termination criteria is much simpler.
When
restricted to a setting where all nodes are active in the first round our algorithm
outputs the diameter in the same time bound. Even though we are more
general, we have significant improvements in message and space complexity.

Related to the computation of the radius and the eccentricity is finding the
network center. To the best of our knowledge, within similar bounds for
space and processing complexity per round, the fastest algorithm so far to
find network center nodes was proposed by Korach, Rotem and
Santoro~\cite{DBLP:journals/toplas/KorachRS84}. This algorithm builds on a
simpler algorithm to find a center of a tree, and the observation that the
center of a general network must also be the center of its own BFS tree: the
initiator node triggers BFS generation at all nodes, picks a center among
all candidates which are centers of their own BFS, and later disseminates
this information to all nodes. The closest center node $c$ to the initiator
node $i$ will receive the confirmation that it is indeed a center of the
network by round $4 \radius(G) + \dist(i,c) + 1$, with total message
complexity of $\Theta(\setsize V \setsize E \log \setsize V)$ bits. Notice
that, in contrast to this, our algorithm is fully symmetrical, not requiring
a distinguished node as initiator (which would have to be chosen in some
way, e.g. by a leader election). Even discounting this factor, our more
general algorithm has no time penalty: in fact, it can be shown that it even
improves this upper bound by at least one round.

%As stated, our focus is time efficiency in the number of rounds. This is
%achieved here under a maximum bandwidth parameter B \cite{Linial:1992p730}
%that allows transmission of at most $n \log n$ bits in each round and edge,
%for a graph $G$ whith $n=\left| V \right|$ nodes and $m=\left| E \right|$
%edges. As in \cite{Peleg:1990p675} the number of messages transmitted in
%our algorithms is at most $O(\diam(G) m)$ since messages can be transmitted
%in all edges until termination in $O(\diam(G))$ rounds.

A more general problem than computing the eccentricities, radius and
diameter is that of computing the distance matrix or all-pairs shortest path
matrix in a network. In fact, relying on this connection, a faster
distributed variant of our algorithm could be designed to compute the radius
and diameter: instead of propagating just the distances, each node
propagates sets of neighbors -- at most by round $2 \diam(G) + 2$ all nodes
would be able to determine the full topology of the network and run a
standard unweighted all-pairs shortest path algorithm with computation
complexity $O(\setsize V \setsize E)$ at the last round. Unfortunately, even
assuming that computation complexity per round is negligible, this algorithm
requires $\Theta({\setsize V}^2)$ space complexity at each node, which is
impractical for large networks. Moreover the message complexity increases to
$\Theta({\setsize E}^2 \log \setsize V)$ bits, and requires a bandwidth of 
$O(\setsize E \log \setsize V)$ bits per link.

With such large bandwidth it is possible to decrease the message complexity
using more elaborated approaches. For example, Kanchi and
Vineyard~\cite{KanchiVineyard04} propose a distributed algorithm to compute
the all-pairs shortest path matrix with message complexity of $O(\setsize V \setsize E \log
\setsize V)$ bits: a spanning tree is first computed using the algorithm
proposed by Awerbuch~\cite{DBLP:conf/stoc/Awerbuch87} and later reused to
propagate the topological information to a root node that computes the
distance matrix and disseminates it to all nodes. The tradeoff for this
optimization is, unfortunately, a substantial increase in
the number of rounds, although still $O(\setsize V)$.

%Exact, slower but possibility less message complexity? %\cite{Magnien:2009p551}

%Non-exact but less space and less message complexity? %\cite{Cardoso:2009p1477} 

%Non-exact, slower but less message complexity. %\cite{GuCheng}

Message complexity can also be reduced by trading off accuracy. For
example, Gu and Cheng~\cite{GuCheng} propose a distributed algorithm
to compute an estimate $\hat{\diam(G)}$ for the diameter, such that
$\diam(G) \leq \hat{\diam(G)} \leq \diam(G)+2$. 
%The message complexity
%for this algorithm in synchronous networks is $O(\setsize V ^ {2.5})$,
%a non-negligible saving in dense networks. 
Unfortunately, the
bandwidth requirements are slightly worse than ours and the
improvement in message complexity does not extend to time complexity: although the authors
do not quantify this measure, it is clear from the algorithm
presentation that the number of rounds until completion is substantially
larger than ours. Moreover, this algorithm also requires a
distinguished node as initiator.

\section{Conclusions}
\label{sec:conclusions}

In this paper we propose a time efficient algorithm that computes in all
nodes of a network the values of the node eccentricity, and the network
diameter and radius. The algorithm is very flexible in the sense that it
does not require a distinguished node, one or more nodes can initiate the
computation, and concurrent initiations have no detrimental impacts on the
various bounds and complexities. 

Under the same communication, space, and computation complexity, the
presented algorithm significantly improves existing time bounds for the
diameter and radius computation. It also slightly improves the time
bounds for the special case of finding center nodes, while relaxing the need
for a special initiator. 

The key to the improvement was to abandon the traditional modular approach.
Instead, our algorithm relies on a very early propagation and aggregation of
the maximum and minimum candidate eccentricities, even before these values
have stabilized.  Together with adequate convergence detection criteria,
this allowed a simple and fast approach to the computation of these
distances.

%\newpage

\bibliography{diam}

\end{document}